\documentclass[english,aps,twocolumn,prl]{revtex4}
\usepackage[T1]{fontenc}
\usepackage[latin9]{inputenc}
\usepackage{amsbsy}
\usepackage{graphicx}
\usepackage{hyperref}
\hypersetup{
    colorlinks=true,
}

\makeatletter
\@ifundefined{textcolor}{}
{%
 \definecolor{BLACK}{gray}{0}
 \definecolor{WHITE}{gray}{1}
 \definecolor{RED}{rgb}{1,0,0}
 \definecolor{GREEN}{rgb}{0,1,0}
 \definecolor{BLUE}{rgb}{0,0,1}
 \definecolor{CYAN}{cmyk}{1,0,0,0}
 \definecolor{MAGENTA}{cmyk}{0,1,0,0}
 \definecolor{YELLOW}{cmyk}{0,0,1,0}
}

\makeatother

\usepackage{babel}
\begin{document}

\preprint{\pagebreak{}This line only printed with preprint option}

\title{Landscape of an exact energy functional}

\author{Aron J. Cohen}

\affiliation{Department of Chemistry, Lensfield Rd, University of Cambridge, Cambridge,
CB2 1EW, UK}

\author{Paula Mori-Sánchez}

\affiliation{Departamento de Química and Instituto de Física de la Materia Condensada
(IFIMAC), Universidad Autónoma de Madrid, 28049, Madrid, Spain}
\begin{abstract}
One of the great challenges of electronic structure theory is the
quest for the exact functional of density functional theory. Its existence
is proven, but it is a complicated multivariable functional that is
almost impossible to conceptualize. In this paper, the asymmetric
two-site Hubbard model is studied, which has a two-dimensional universe
of density matrices. The exact functional becomes a simple function
of two variables whose three dimensional energy landscape can be visualized
and explored. A walk on this unique landscape, tilted to an angle
defined by the one-electron Hamiltonian, gives a valley whose minimum
is the exact total energy. This is contrasted with the landscape of
some approximate functionals, explaining their failure for electron
transfer in the strongly correlated limit. We show concrete examples
of pure-state density matrices that are not $v$-representable due
to the underlying non-convex nature of the energy landscape. For the
first time, the exact functional is calculated for all numbers of
electrons, including fractional, allowing the derivative discontinuity
to be visualized and understood. The fundamental gap for all possible
systems is obtained solely from the derivatives of the exact functional. 
\end{abstract}
\maketitle

In 1964 Hohenberg and Kohn \cite{Hohenberg64864} established density
functional theory (DFT) showing that the electron density, $\rho$,
is all that is is necessary to determine the exact energy of many
electron systems. However, all the challenge of electronic structure
is then moved into an unknown universal functional of the density,
$\mathcal{F}[\rho]$. For a wavefunction, $\Psi_{v}$, that is the
ground-state solution of the Schrödinger equation with potential $v$,
\begin{equation}
E_{v}=\min_{\Psi}\langle\Psi|\hat{H}|\Psi\rangle=\langle\Psi_{v}|T+V_{ee}|\Psi_{v}\rangle+{\rm Tr}({\bf \boldsymbol{\rho_{v}v}})
\end{equation}
simply subtracting off the one-electron term, gives the exact Hohenberg-Kohn
functional for $\rho_{v}$ ($\Psi_{v}\rightarrow\rho_{v}$) 
\begin{equation}
\mathcal{F}^{{\rm HK}}[\rho_{v}]=E_{v}-{\rm Tr}({\bf \boldsymbol{\rho_{v}{\it v}}})=\langle\Psi_{v}|T+V_{ee}|\Psi_{v}\rangle.\label{eq:FHK}
\end{equation}
This procedure can be carried out for many different $v$, to obtain
many points of the exact functional $\mathcal{F}^{{\rm HK}}[\rho_{v}]$.
A question arises of whether all possible densities are achievable.
This is the problem of $v$-representability, that is addressed by
the constrained search by Levy and Lieb \cite{Levy796062,Lieb83243}
following earlier work by Percus \cite{Percus7889}
\begin{equation}
\mathcal{F}^{{\rm Levy}}[\rho]=\min_{\Psi\rightarrow\rho}\langle\Psi|T+V_{ee}|\Psi\rangle.
\end{equation}
This functional is defined for all possible densities coming from
a $N$-electron wavefunction, including those that are not obtainable
as the ground-state solution of a Schrödinger equation (not $v$-representable).
Once the exact functional is known, the total energy is obtained by
minimization only over densities, 
\begin{equation}
E_{v}[\rho]=\min_{\rho}\left\{ \mathcal{F}[\rho]+{\rm Tr}(\boldsymbol{\rho v)}\right\} .
\end{equation}
The exact functional of the first-order density matrix, $\gamma$,
can be derived \cite{Gilbert752111,Levy796062} 
\begin{eqnarray}
F^{{\rm Levy}}[\gamma] & = & \min_{\Psi\rightarrow\gamma}\langle\Psi|V_{ee}|\Psi\rangle,\label{eq:Levygamma}
\end{eqnarray}
and used similarly, where the kinetic energy term is now a known linear
functional of $\gamma$ 
\begin{equation}
E_{v}[\rho]=\min_{\gamma}\left\{ F[\gamma]+{\rm Tr}({\bf T}\boldsymbol{\gamma})+{\rm Tr}(\boldsymbol{v\gamma)}\right\} .
\end{equation}

\begin{figure*}
\includegraphics[angle=-90,width=0.9\textwidth]{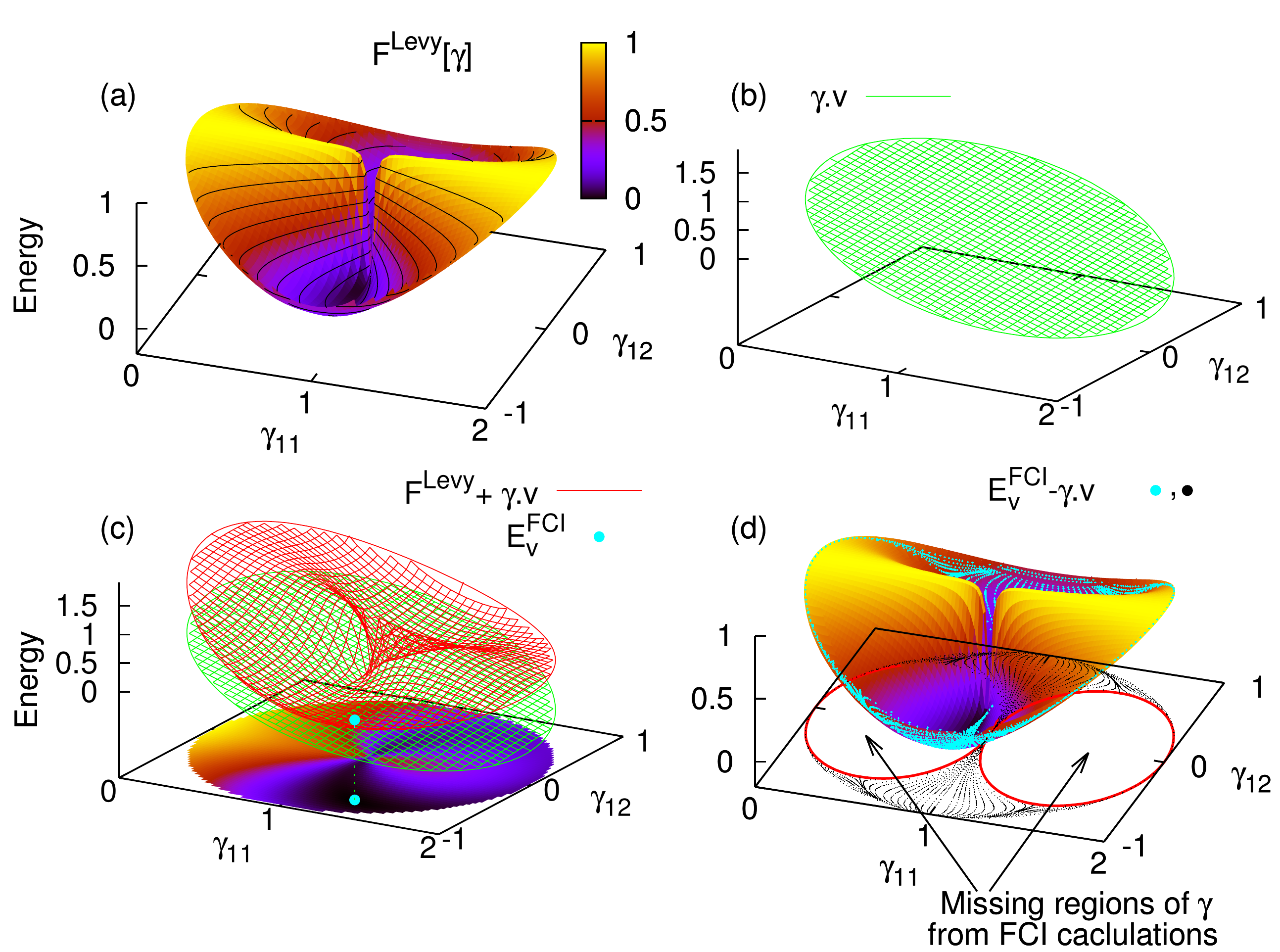}\caption{Energy landscape of the exact functional (a) $F^{{\rm Levy}}[\gamma]$
for all allowable density matrices of the two site Hubbard model.
(b) The one electron term, $\gamma.v$, for $t=0.1$ and $\Delta\epsilon=0.9$,
which is purely a flat plane. (c) Illustration of the minimization
of the exact functional adding on the same $\gamma.v$ term to give
the FCI energy and density matrix, $\left\{ E_{v}^{{\rm FCI}},\gamma_{v}^{{\rm FCI}}\right\} $
. (d) $F^{{\rm Levy}}[\gamma]$ and 6552 points of $\left\{ F^{{\rm HK}}[\gamma_{v}^{{\rm FCI}}],\gamma_{v}^{{\rm FCI}}\right\} $
that show the $E_{v}^{{\rm FCI}}$ subtracting the one electron term
(Eq. \ref{eq:HK functional}) at $\gamma_{v}^{{\rm FCI}}$ for many
different $v$. \label{fig:Four views F[rho]}}
\end{figure*}

In this Letter, the nature of the exact first-order density matrix
functional is revealed by considering the asymmetric two-site Hubbard
model. In this universe, the fundamental equations are tractable and
the exact functional becomes a visualizable three dimensional energy
landscape in the space of density matrices. We demonstrate how this
one universal landscape gives the exact energy of all possible systems,
for all numbers of electrons including fractional. This connected
view of the functional for all density matrices makes clear the reasons
for the failure of approximate functionals, and allows us to answer
the questions of whether there are density-matrices which are not
$v$-representable and also how the derivatives of the exact functional
give the fundamental gap.

The asymmetric two-site Hubbard \cite{Hubbard26111963} model describes
interacting electrons on a lattice of two sites that contains the
physics of electron transfer and has even recently been experimentally
described using two ultracold fermionic atoms \cite{Murmann15080402}.
It has the Hamiltonian
\begin{equation}
\hat{H}=-t\sum_{\sigma}\left(c_{1\sigma}^{\dagger}c_{2\sigma}+c_{2\sigma}^{\dagger}c_{1\sigma}\right)+U\sum_{i}\hat{n}_{i\alpha}\hat{n}_{i\beta}+\sum_{i\sigma}\epsilon_{i}\hat{n}_{i\sigma}\label{eq:Hubbard model}
\end{equation}
where the site index $i=1,2$, spin index $\sigma=\alpha,\beta$ and
the number operator is $\hat{n}_{i\sigma}=c_{i\sigma}^{\dagger}c_{i\sigma}$.
There has been recent work on the exact functional in this model from
Fuks \emph{et al} \cite{Fuks14062512,Fuks1414504}, Carrascal \emph{et
al} \cite{Carrascal15393001}, Pastor and coworkers \cite{LopezSandoval02155118,Saubanere1103511},
Requist \emph{et al} \cite{Requist08235121}, and in other systems
\cite{Teale10164115,Kvaal14518,Wagner14045109}. 

\begin{figure*}[t]
\includegraphics[angle=-90,width=0.9\textwidth]{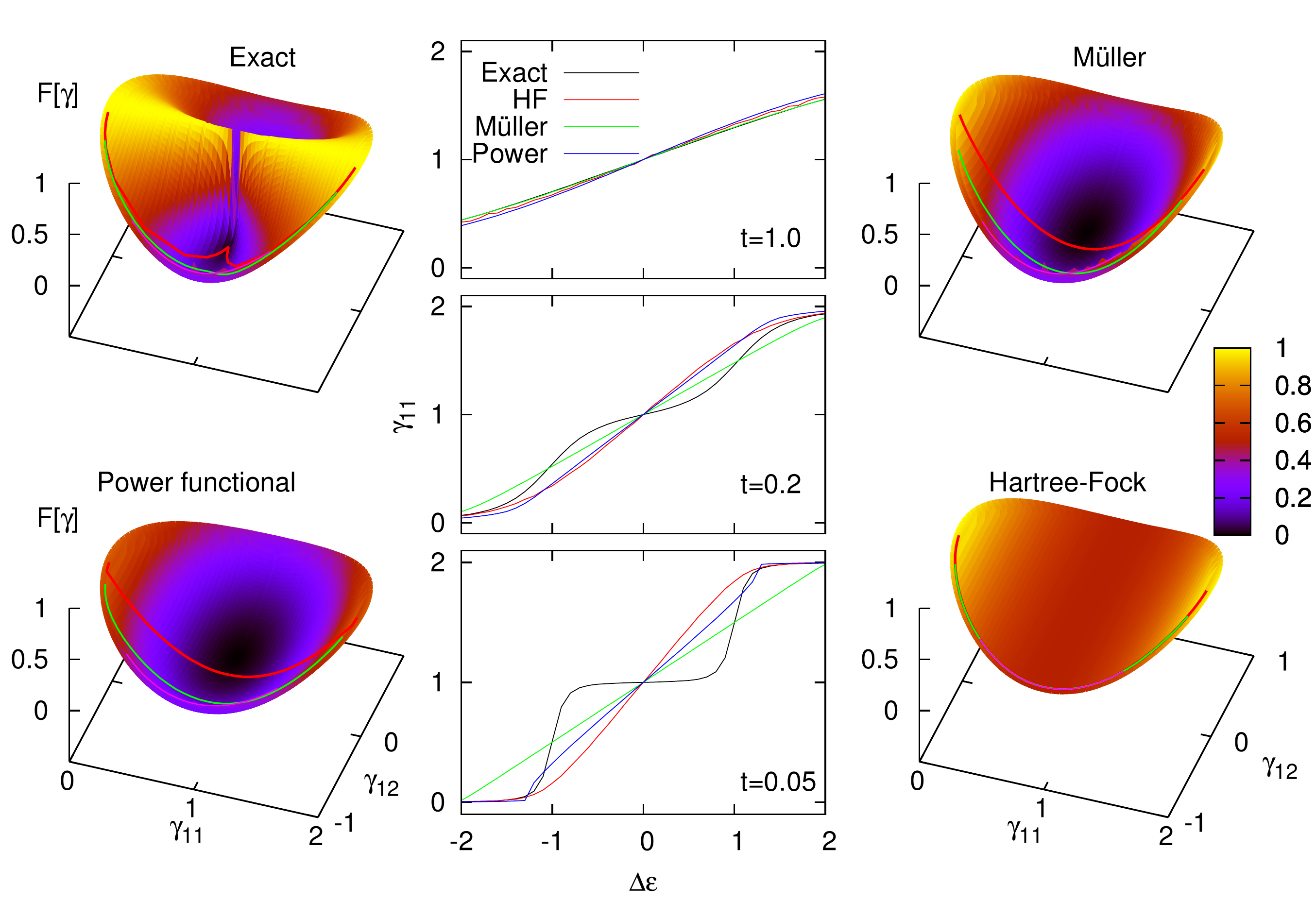}\caption{Entire landscape of $F[\gamma]$ for the exact functional, and three
approximate density-matrix functionals (see supplementary information
for more details). The minimizing values $\left\{ F[\gamma_{v}],\gamma_{v}\right\} $
for three lines of $v$ ($-2\le\Delta\epsilon<2$) with $t=1,0.2,0.05$
are plotted on the surfaces in purple, green and red, respectively\label{fig:strong correlation}.
The central plots show the failure of approximate functionals to correctly
describe the electron transfer ($\gamma_{11}$ vs $\Delta\epsilon)$
as $t$  approaches the strongly correlated limit (see supplementary
animations). }
\end{figure*}

The parameters that define a particular model are the hopping between
the sites, $t$, on-site energies $\epsilon_{1}/\epsilon_{2}$ and
the electron-electron repulsion penalty due to double occupation of
a site, $U$. The physics is completely determined by $\Delta\epsilon=\epsilon_{1}-\epsilon_{2}$
and the ratio $U/t$, therefore, in this work $U$ is fixed at 1 and
$t$ and $\Delta\epsilon$ are the chosen variables. The kinetic and
on-site potential part of the Hamiltonian, which together we denote
as $v$, is a real symmetric 2x2 matrix defined by parameters $t$
and $\Delta\epsilon$ 
\begin{equation}
v=\left(\begin{array}{cc}
\Delta\epsilon/2 & -t\\
-t & -\Delta\epsilon/2
\end{array}\right)
\end{equation}
and the 2x2 density matrix$,$ $\gamma_{ij}=\sum_{\sigma}\langle\Psi|c_{i\sigma}^{\dagger}c_{j\sigma}|\Psi\rangle$
is
\begin{equation}
\gamma=\left(\begin{array}{cc}
\gamma_{11} & \gamma_{12}\\
\gamma_{12}^{*} & (2-\gamma_{11})
\end{array}\right)
\end{equation}
leading to a total energy for real density matrices 
\begin{eqnarray}
E_{v} & = & -2\gamma_{12}t+\gamma_{11}\Delta\epsilon/2-(2-\gamma_{11})\Delta\epsilon/2+F[\gamma].
\end{eqnarray}
The exact functional can be obtained and understood from different
perspectives. Firstly, for any $\gamma$ that comes from an exact
diagonalization full configuration interaction (FCI) calculation with
one-electron Hamiltonian $v$, the Hohenberg-Kohn functional is given
by
\begin{equation}
F^{{\rm HK}}[\gamma_{v}]=E_{v}^{{\rm FCI}}+2\gamma_{12}t-\gamma_{11}\Delta\epsilon/2+(2-\gamma_{11})\Delta\epsilon/2.\label{eq:HK functional}
\end{equation}
The second way is the constrained search over real singlet wavefunctions
\begin{eqnarray}
\Psi & = & \frac{a}{\sqrt{2}}\left[\mathcal{A}(\phi_{1}\alpha\phi_{2}\beta)+\mathcal{A}(\phi_{2}\alpha\phi_{1}\beta)\right]\nonumber \\
 &  & +b\mathcal{A}(\phi_{1}\alpha\phi_{1}\beta)+c\mathcal{A}(\phi_{2}\alpha\phi_{2}\beta)\label{eq:Wavefunction}
\end{eqnarray}
which can be simplified to an expression (see Refs. \cite{Saubanere1103511,Carrascal15393001}
and supplementary information (SI) for more details) 
\begin{equation}
F^{{\rm Levy}}[\gamma]=\frac{\gamma_{12}^{2}\left(1-\sqrt{1-\gamma_{12}^{2}-[\gamma_{11}-1]^{2}}\right)+2[\gamma_{11}-1]^{2}}{2\left(\gamma_{12}^{2}+[\gamma_{11}-1]^{2}\right)}.\label{eq:HubbardLevy}
\end{equation}
Thirdly, it can be viewed as the exact functional in density matrix
functional theory for two electrons. From the work of Löwdin and Shull
in 1956 \cite{Lowdin561730} using the natural orbitals $|a\rangle$
and $|b\rangle$ ($|p\rangle=\sum_{i=1,2}C_{pi}c_{i}^{\dagger}|{\rm vac}\rangle$)
and their occupation numbers $n_{a}$ and $n_{b}$ that diagonalize
$\gamma$, it can be derived that 
\begin{equation}
F^{{\rm LS}}[\gamma]=\frac{1}{2}n_{a}\langle aa|aa\rangle+\frac{1}{2}n_{b}\langle bb|bb\rangle-\sqrt{n_{a}n_{b}}\langle aa|bb\rangle\label{eq:Lowdin}
\end{equation}
where the two-electron integral is $\langle pp|qq\rangle=U\sum_{i=1,2}C_{pi}^{2}C_{qi}^{2}$.
This gives exact agreement with the constrained search expression,
Eq. (\ref{eq:HubbardLevy}) and has been utilized in functionals such
as the AGP natural orbital functional \cite{barbiellini_natural_2000,barbiellini_treatment_2001}
and PNOF5 \cite{Piris11164102} (see SI). There are two further possible
routes to the exact functional (details in the SI): the extension
over pure-state wavefunctions to complex, and the Lieb maximization\cite{Lieb83243},
$F^{{\rm Lieb}}[\gamma]=\sup_{v}\left\{ E_{v}-\gamma.v\right\} $. 

$F^{{\rm Levy}}[\gamma]$ is shown in Fig. 1a. for the allowable density
matrices $\left(\gamma_{11}-1\right)^{2}+\gamma_{12}^{2}\le1$. It
is represented as a unique surface of hills and a valley in a bowl
type shape, with a channel through the centre (at $\gamma_{11}=1$)
and hills on both sides (reaching 1 at $\gamma_{12}=0$). This defines
the energy landscape that maps every possible system to its corresponding
exact energy. 

The exact functional is an energy landscape with only one minimum,
so how does it give rise to all possible FCI energies? This can be
pictured in a very physical manner by considering a walk on this landscape,
placed upon a flat surface tilted to the angle given by the one-electron
potential, which gives a valley whose minimum equals exactly the FCI
solution. Fig. 1b shows the one electron term for a particular $v$,
defined by $t=0.1$ and $\Delta\epsilon=0.9$, and Fig. 1c shows the
addition of this with the exact functional, $F^{{\rm Levy}}[\gamma]+\gamma.v$,
whose minimum is at the FCI energy, $E_{v}^{{\rm FCI}}$, and FCI
density matrix, $\gamma_{v}$. This holds for every possible $v$.
Thus, once the exact functional is known, it gives the exact solution
of any system by means of an almost trivial calculation.

We have performed a large number of FCI calculations varying the two
free parameters, $-10<t<10$ and $-10<\Delta\epsilon<10$. Fig. 1d
illustrates the result of over 6000 FCI calculations subtracting off
the one electron term, $\gamma.v$, to give the $F^{{\rm HK}}[\gamma]$
of Eq. (\ref{eq:HK functional}). Every single light blue dot, representing
many $F^{{\rm HK}}[\gamma_{v}]$, lies on the surface of $F^{{\rm Levy}}[\gamma]$.
However, the one-particle density matrices $\gamma_{v}$ that result
from all these FCI calculations cover only a small fraction of the
space (seen as the black dots projected onto the base of the plot
with more details in SI). The rest of the density matrices are not
$v$-representable, even though they are $N$-representable. From
the perspective of the exact functional, it is clear why these density
matrices can never be found, as they correspond to the hills of the
surface where the $F^{{\rm Levy}}[\gamma]$ lies inside a convex containing
surface (see SI). Addition of the one electron interaction term, which
is purely linear in the variables $\gamma_{11}$ and $\gamma_{12}$,
as pictured in Fig. 1b, means that these points can never be minima,
and hence cannot be a FCI solution. In terms of the functional it
corresponds to where the second derivatives of the functional are
no longer positive definite as seen by the a negative lowest eigenvalue
of the Hessian matrix of second derivatives, $H_{ij}=\frac{\partial^{2}F}{\partial\gamma_{1i}\partial\gamma_{1j}}$,
(see SI). It should also be noted that the lowest energy wavefunctions
of the non-$v$-representable density matrices cannot be written in
a Gutzwiller form \cite{Gutzwiller63159} (see SI). The non-$v$-representable
region highlights the key distinction between $F^{{\rm Levy}}[\gamma]$
derived from pure-state wavefunctions, which can be concave, versus
the $F^{{\rm Lieb}}[\gamma]$ functional derived from ensembles by
a Legendre-Fenchel transform, which is proven to be everywhere convex
\cite{Lieb83243}.

The derivatives of the functional (expressions in SI) satisfy the
Euler equation and give the one-electron Hamiltonian needed, 
\begin{equation}
\frac{\partial F[\gamma]}{\partial\gamma}=-v+C.\label{eq:Euler equation}
\end{equation}

\begin{figure}
\includegraphics[angle=-90,width=1\columnwidth]{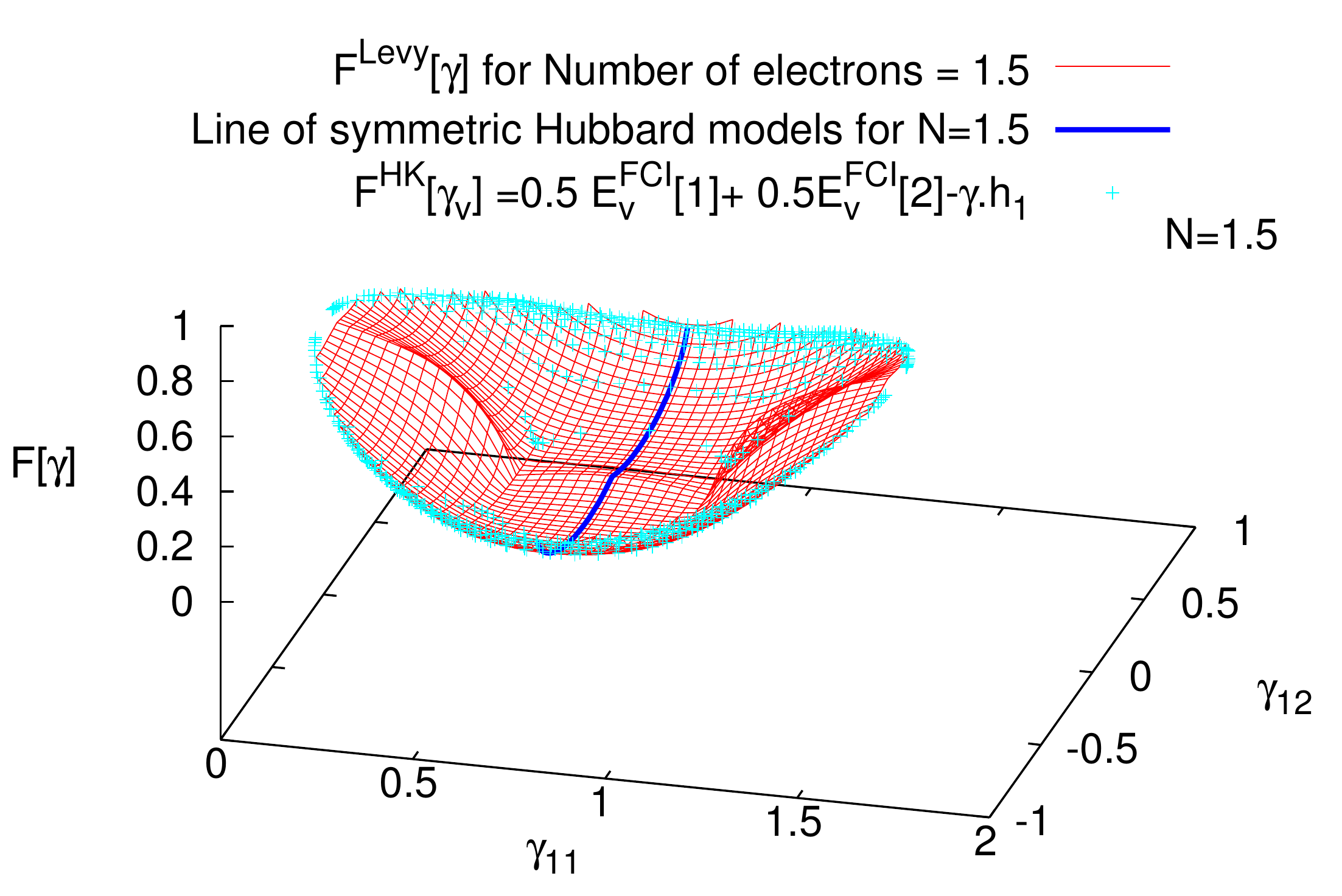}

\caption{The exact functionals of Eqs (\ref{eq:FPPLB}) and (\ref{eq:FHKfrac})
for $N=1.5$ electrons, which gives back the exact energy of every
system with 1.5 electrons.\label{fig:F[N=00003D1.5]}}
\end{figure}

Now, consider the physics of electron transfer, by varying $\Delta\epsilon$,
from the weakly correlated ($U/t=1$) to strongly correlated ($U/t=20$)
regimes as depicted in Fig. \ref{fig:strong correlation}. Correctly
describing this electron transfer in the strongly correlated regimes
is one of the great challenges of electronic structure, as demonstrated
in Fig. \ref{fig:strong correlation} by the failure of approximate
density matrix functionals such as Müller \cite{Muller84446} and
Power functionals \cite{Sharma08201103}. The approximate functionals
do not correctly describe the entire landscape and thus completely
fail to describe electron transfer (see animation in SI). This is
related to the complete failure of all currently used density functionals
for the electron transfer in a two-electron molecular type challenge
(see HZ$^{\{2e\}}$of Ref. \cite{Mori-Sanchez1414378}).

\begin{figure}
\includegraphics[angle=-90,width=1\columnwidth]{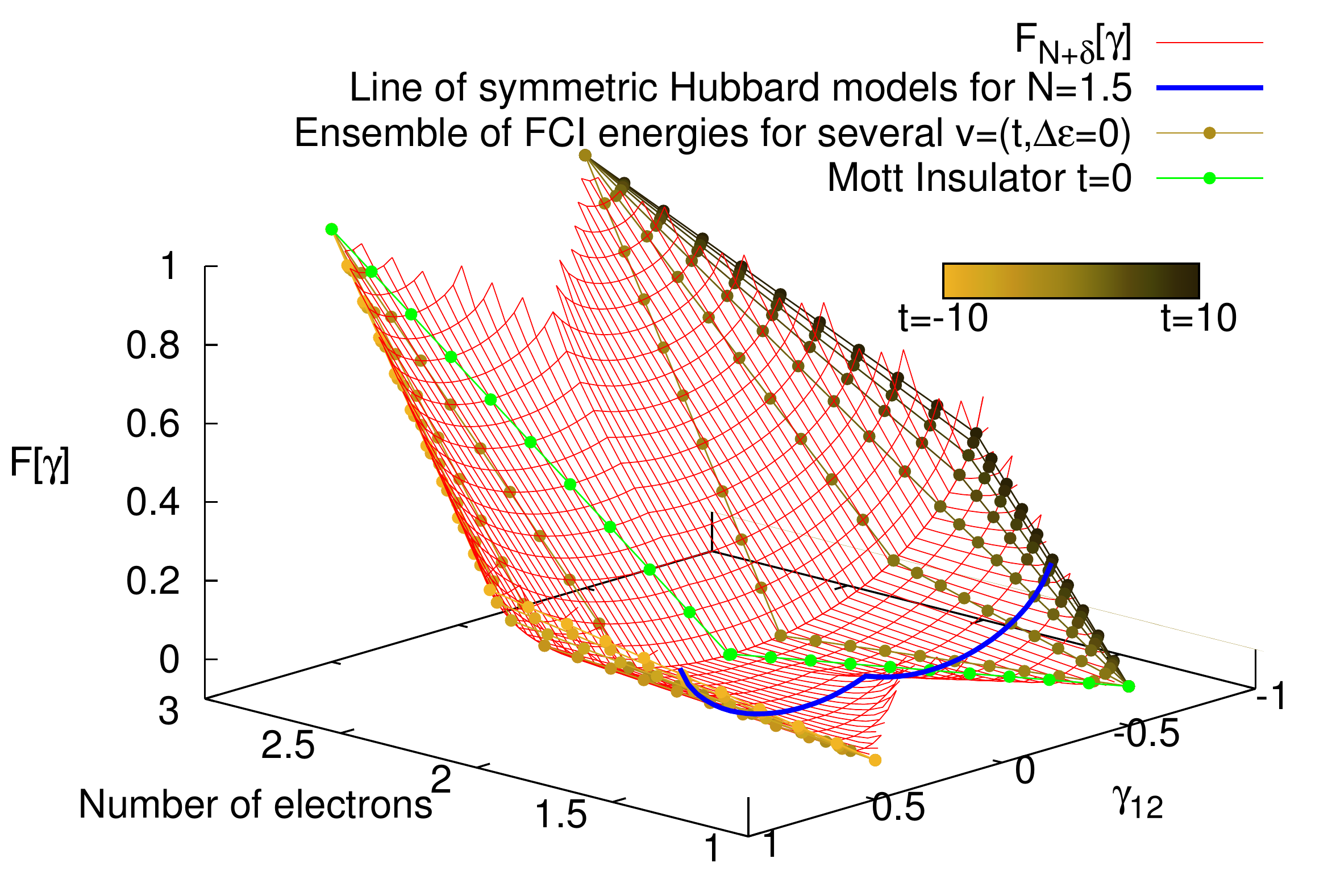}\caption{The exact functionals of Eqs. (\ref{eq:FPPLB}) and (\ref{eq:FHKfrac})
for all numbers of electrons ($1\le N\le3)$ in the symmetric two
site Hubbard model.\label{fig:Fractional Symmetric F}}
\end{figure}

The exact functional can be calculated for all numbers of electrons
($0\le N\le4)$; the integer parts are trivial and given in the SI.
For non-integer numbers of electrons, the functional is constructed
using the Perdew, Parr, Levy and Balduz (PPLB)\cite{Perdew821691}
ensemble extension to search over many-electron density matrices 
\begin{eqnarray}
\Gamma_{N+\delta} & = & c_{0}|\Psi_{0}\rangle\langle\Psi_{0}|+c_{1}|\Psi_{1}\rangle\langle\Psi_{1}|+c_{2}|\Psi_{2}\rangle\langle\Psi_{2}|\nonumber \\
 &  & +c_{3}|\Psi_{3}\rangle\langle\Psi_{3}|+c_{4}|\Psi_{4}\rangle\langle\Psi_{4}|\label{eq:WavefunctionEnsemble}
\end{eqnarray}
with $\sum_{i}c_{i}=1$ and $\sum_{i}c_{i}.i=N+\delta\ (0\le\delta\le1)$.
Thus, we explicitly construct the fractional extension 
\begin{equation}
F_{N+\delta}[\gamma]=\min_{\Gamma_{N+\delta}\rightarrow\gamma}{\rm Tr}[\Gamma_{N+\delta}V_{ee}],\label{eq:FPPLB}
\end{equation}
where, unlike PPLB, we have not assumed convexity of the energy versus
$N$. That is, rather than using $\Gamma_{N+\delta}=c_{N}|\Psi_{N}\rangle\langle\Psi_{N}|+c_{N+1}|\Psi_{N+1}\rangle\langle\Psi_{N+1}|$,
we explicitly search over ensembles of all $N$-electron wavefunctions
($N=0,1,2,3,$ and $4$) as in Eq. (\ref{eq:WavefunctionEnsemble})
(see SI).

Fig. \ref{fig:F[N=00003D1.5]} shows the extension of the exact functional
to fractional numbers of electrons for $N+\delta=1.5$. We obtained
$F_{N+\delta}[\gamma]$ for all the possible density matrices, where
the minimum is actually given only by the combination of $N$ and
$N+1$ (see supplementary information for more details). We also find
that all the appropriate ensembles of FCI energies subtracting off
the one electron term using the ensemble of density matrices, 
\begin{eqnarray}
F_{N+\delta}^{{\rm HK}}[v] & = & (1-\delta)E_{v}^{{\rm FCI}}[N]+\delta E_{v}^{{\rm FCI}}[N+1]\nonumber \\
 &  & -\left[(1-\delta)\gamma_{v}^{N}+\delta\gamma_{v}^{N+1}\right].v\label{eq:FHKfrac}
\end{eqnarray}
lie perfectly on the functional surface for all values of $v$ and
$\delta$. Additionally, just like for integer electrons, a walk on
this surface tilted to the angle of any one-electron potential (analogously
to Fig. 1c) gives a minimum point that exactly agrees with the ensemble
FCI energy.

The knowledge of the exact functional for fractional numbers of electrons
connects to the band-gap problem. This is the question of whether
the fundamental gap, defined as the difference of the ionization energy
and electron affinity, can be given by the derivatives of the exact
functional. For simplicity, consider only the symmetric Hubbard dimer
with different numbers of electrons. In Fig. \ref{fig:Fractional Symmetric F}
the exact functional is shown for $1\le N\le3$, along with several
points of the ensemble $F_{N+\delta}^{{\rm HK}}[v]$ with $v=\{-1<t<1,\Delta\epsilon=0\}$
(see also animations in SI). For every $v$, $F_{N+\delta}^{{\rm HK}}[v]$
traces out a straight line versus particle number with a clear derivative
discontinuity at $N=2$, hence the derivatives of the exact functional
give the contribution to the fundamental gap 
\begin{equation}
\left.\frac{\partial F[\gamma]}{\partial N_{+}}\right|_{v}-\left.\frac{\partial F[\gamma]}{\partial N_{-}}\right|_{v}=F[\gamma^{N+1}]+F[\gamma^{N-1}]-2F[\gamma^{N}].\label{eq:FracDeriv}
\end{equation}
If there is no discontinuity in the density matrix, which is the case
of a Mott insulator, the entirety of the fundamental gap is given
by the exact functional (Eq. \ref{eq:FracDeriv}). This is illustrated
as the green line in Fig. 4 for the symmetric Hubbard model with $t=0$
and $1\le N\le3$, and has a direct correspondence to the gap of infinitely
stretched H$_{2}$ \cite{Mori-Sanchez09}. Nevertheless, most systems
have a discontinuity in the density matrix, $\gamma^{N+1}-\gamma^{N}\ne\gamma^{N}-\gamma^{N-1}$,
giving rise to a discontinuous derivative even for the one electron
term, which is an entirely smooth flat plane. However, the direction
in which $\gamma$ changes upon electron addition or removal is already
determined by derivatives of $F$ whilst keeping the derivative in
the direction of fixed $N$ to be constant
\begin{equation}
\gamma_{N\pm1}=\gamma_{N}+\left.\frac{\delta\gamma}{\delta N_{\pm}}\right|_{\left.\frac{\partial F}{\partial\gamma}\right|_{N}}.
\end{equation}
Hence, the fundamental gap is solely determined by the derivatives
of the functional itself,
\begin{eqnarray*}
{\rm Gap[\gamma_{N}]} & = & \left(\left.\frac{\partial F[\gamma]}{\partial N_{+}}\right|_{\left.\frac{\partial F[\gamma]}{\partial\gamma}\right|_{N}}-\left.\frac{\partial\gamma}{\partial N_{+}}\right|_{\left.\frac{\partial F}{\partial\gamma}\right|_{N}}.\left.\frac{\partial F[\gamma]}{\partial\gamma}\right|_{N}\right)\\
 &  & -\left(\left.\frac{\partial F[\gamma]}{\partial N_{-}}\right|_{\left.\frac{\partial F[\gamma]}{\partial\gamma}\right|_{N}}-\left.\frac{\partial\gamma}{\partial N_{-}}\right|_{\left.\frac{\partial F}{\partial\gamma}\right|_{N}}.\left.\frac{\partial F[\gamma]}{\partial\gamma}\right|_{N}\right)
\end{eqnarray*}

Overall, it is amazing to have a universe that turns any question
about the exact functional into simple movements of a three-dimensional
energy landscape. Walks on this landscape and its valley and hills
correspond to important physical concepts such as the exact energies
of every possible system and domains of non-$v$-representable density
matrices. Furthermore, in the direction of changing particle number
there is a continuous surface that has a derivative discontinuity
at the integers, giving all possible fundamental gaps, including Mott
insulators. The whole landscape of the exact functional is itself
an infinite number of exact constraints, such that any approximation
must approach and be mathematically proximal to it for the entire
universe. It is this connected view of the exact functional for a
family of densities in a global landscape that truly highlights a
path for the improvement of approximate functionals. 
\begin{acknowledgments}
We gratefully acknowledge funding from Ramon y Cajal (PMS) and the
Royal Society (AJC). PMS also acknowledges grant FIS2012-37549 from
the Spanish Ministry of Science.
\end{acknowledgments}
\bibliographystyle{apsrev4-1}

%

\setcounter{equation}{0}
\setcounter{figure}{0}
\renewcommand{\theequation}{S\arabic{equation}}
\renewcommand{\thefigure}{S\arabic{figure}}

%
%
%
%
%
%
%
%
%
%
%
%
%
%
%
%

\begin{widetext}
\begin{center}
{\Large \bf Supplementary Information to "Landscape of an exact functional"}\\
\vspace{0.5cm}
Aron J. Cohen \\
{\it Department of Chemistry, Lensfield Rd, University of Cambridge, Cambridge, CB2 1EW, UK}\\
\vspace{0.5cm}
Paula Mori-S\'anchez\\
\noindent
{\it Departamento de Química and Instituto de Física de la Materia Condensada
(IFIMAC), Universidad Autónoma de Madrid, 28049, Madrid, Spain}
\end{center}
\vspace{0.5cm}

\end{widetext}

\subsection{Derivation of Eq. (13)}

To derive the exact functional, 
\begin{equation}
F^{{\rm Levy}}=\min_{\Psi\rightarrow\gamma}\langle\Psi|V_{ee}|\Psi\rangle
\end{equation}
consider the minimization over real singlet wavefunctions

\begin{eqnarray}
\Psi & = & \frac{a}{\sqrt{2}}\left[\mathcal{A}(\phi_{1}\alpha\phi_{2}\beta)+\mathcal{A}(\phi_{2}\alpha\phi_{1}\beta)\right]\nonumber \\
 &  & +b\mathcal{A}(\phi_{1}\alpha\phi_{1}\beta)+c\mathcal{A}(\phi_{2}\alpha\phi_{2}\beta).
\end{eqnarray}
in terms of the parameters $a,b$ and $c$ along with the normalization
$a^{2}+b^{2}+c^{2}=1$ and the elements of the density-matrix $\gamma_{ij}=\sum_{\sigma}\langle\Psi|c_{i\sigma}^{\dagger}c_{j\sigma}|\Psi\rangle$
giving $\gamma_{11}=2b^{2}+a^{2}$ and $\gamma_{12}=\sqrt{2}\left(ba+ac\right)$.
The two-electron energy comes only from the $\langle11|11\rangle$
and $\langle22|22\rangle$ integrals, which are $U$, as all other
integrals are 0, so only the second determinant with itself and the
third determinant with itself contribute, giving

\begin{equation}
F[\Psi]=U(b^{2}+c^{2})=U(1-a^{2})
\end{equation}
It is also satisfied that
\begin{equation}
\gamma_{11}-1=b^{2}-c^{2}.\label{eq:b2-c2}
\end{equation}
Therefore, using $\gamma_{12}$ gives
\begin{equation}
(b+c)=\frac{\gamma_{12}}{\sqrt{2}a}\label{eq:b+c}
\end{equation}
and combining with Eq. (\ref{eq:b2-c2}) leads to
\begin{equation}
(b-c)=\frac{\left(\gamma_{11}-1\right)\sqrt{2}a}{\gamma_{12}}\label{eq:b-c}
\end{equation}
Now, square Eqs. (\ref{eq:b+c}) and (\ref{eq:b-c}), to give
\begin{equation}
(b+c)^{2}=\frac{\gamma_{12}^{2}}{2a^{2}}
\end{equation}
and
\begin{equation}
(b-c)^{2}=\frac{\left(\gamma_{11}-1\right)^{2}2a^{2}}{\gamma_{12}^{2}}.
\end{equation}
Adding these two has the result
\begin{eqnarray}
2b^{2}+2c^{2} & = & \frac{\gamma_{12}^{2}}{2a^{2}}+\frac{(\gamma_{11}-1)^{2}2a^{2}}{\gamma_{12}^{2}}.
\end{eqnarray}
Using the normalization, gives
\begin{equation}
(2-2a^{2})=\frac{\gamma_{12}^{2}}{2a^{2}}+\frac{(\gamma_{11}-1)^{2}2a^{2}}{\gamma_{12}^{2}}
\end{equation}
which leads to a quadratic equation for $a^{2}$
\begin{equation}
\frac{[(\gamma_{11}-1)^{2}+\gamma_{12}^{2}]}{\gamma_{12}^{2}}a^{4}-a^{2}+\frac{\gamma_{12}^{2}}{4}=0
\end{equation}
with solution 
\begin{equation}
a^{2}=\frac{\gamma_{12}^{2}\left(1\pm\sqrt{1-(\gamma_{11}-1)^{2}-\gamma_{12}^{2}}\right)}{2[(\gamma_{11}-1)^{2}+\gamma_{12}^{2}]}.
\end{equation}
\begin{figure}[t]
\includegraphics[angle=-90,width=1\columnwidth]{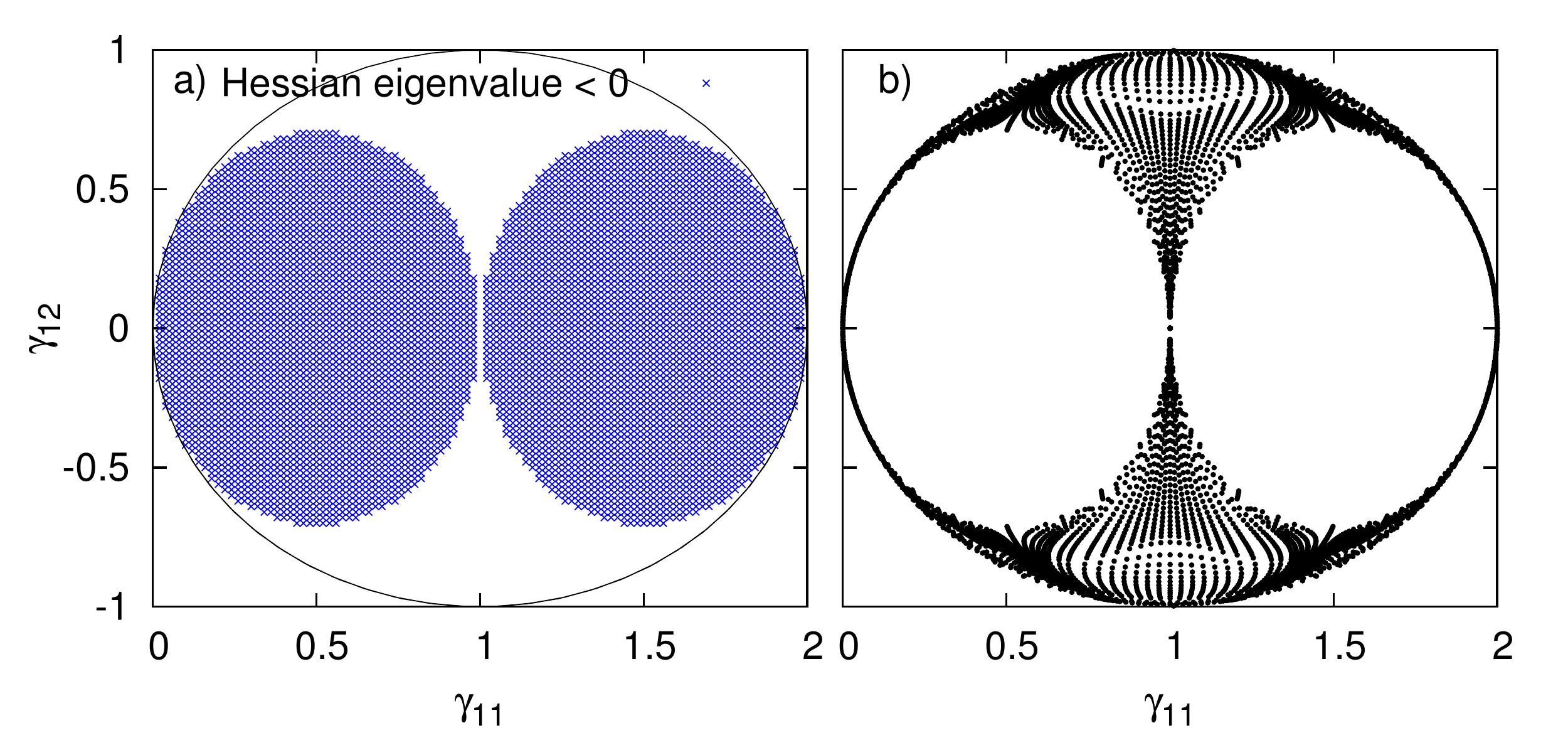}\caption{The plane of all possible density matrices illustrating the non-$v$-representability
of many of the allowable $\gamma$. a) The second derivatives of the
exact functional showing the points where the lowest hessian eigenvalue
is < 0 from Eqs \ref{eq:hess1}-\ref{eq:hess3} and b) the density
matrices, $\gamma$, achieved in 6552 FCI calculations for $-10<t<10$
and $-10<\Delta\epsilon<10.$\label{fig:vrep}}
\end{figure}
Taking the plus combination gives the lowest energy 
\begin{eqnarray}
E & = & 1-a^{2}\nonumber \\
 & = & 1-\frac{\gamma_{12}^{2}\left(1+\sqrt{1-(\gamma_{11}-1)^{2}-\gamma_{12}^{2}}\right)}{2[(\gamma_{11}-1)^{2}+\gamma_{12}^{2}]}\nonumber \\
 & = & \frac{2[(\gamma_{11}-1)^{2}+\gamma_{12}^{2}]-\gamma_{12}^{2}\left(1+\sqrt{1-(\gamma_{11}-1)^{2}-\gamma_{12}^{2}}\right)}{2[(\gamma_{11}-1)^{2}+\gamma_{12}^{2}]}\nonumber \\
 & = & \frac{2(\gamma_{11}-1)^{2}+\gamma_{12}^{2}\left(1-\sqrt{1-(\gamma_{11}-1)^{2}-\gamma_{12}^{2}}\right)}{2[(\gamma_{11}-1)^{2}+\gamma_{12}^{2}]}.
\end{eqnarray}
This agrees with Eq. (13) of the paper.

The derivatives of the exact functional can be evaluated analytically
and are used in Fig. \ref{fig:vrep}. \begin{widetext} {\small{
\[
\frac{\partial E}{\partial\gamma_{11}}=\frac{4(\gamma_{11}-1)+(\gamma_{11}-1)\gamma_{12}^{2}/\sqrt{1-(\gamma_{11}-1)^{2}-\gamma_{12}^{2}}}{2\left((\gamma_{11}-1)^{2}-\gamma_{12}^{2}\right)}-\frac{(\gamma_{11}-1)\left(2(\gamma_{11}-1)^{2}+\gamma_{12}^{2}\left(1-\sqrt{1-(\gamma_{11}-1)^{2}-\gamma_{12}^{2}}\right)\right)}{\left[(\gamma_{11}-1)^{2}+\gamma_{12}^{2}\right]^{2}}
\]
}}{\small \par}

and{\footnotesize{
\[
\frac{\partial E}{\partial\gamma_{12}}=\frac{\gamma_{12}^{3}/\sqrt{1-(\gamma_{11}-1)^{2}-\gamma_{12}^{2}}+2\gamma_{12}\left(1-\sqrt{1-(\gamma_{11}-1)^{2}-\gamma_{12}^{2}}\right)}{2((\gamma_{11}-1)^{2}+\gamma_{12}^{2})}-\frac{2\gamma_{12}(\gamma_{11}-1)^{2}+\gamma_{12}^{3}\sqrt{1-(\gamma_{11}-1)^{2}-\gamma_{12}^{2}}}{\left[(\gamma_{11}-1)^{2}+\gamma_{12}^{2}\right]^{2}}.
\]
}}{\scriptsize{
\begin{eqnarray}
\frac{\partial^{2}E}{\partial^{2}\gamma_{11}^{2}} & = & \frac{-4(\gamma_{11}-1)^{2}\left((\gamma_{11}-1)^{2}+\gamma_{12}^{2}\right)\left(\frac{\gamma_{12}^{2}}{\sqrt{-\gamma_{11}^{2}+2\gamma_{11}-\gamma_{12}^{2}}}+4\right)+8(\gamma_{11}-1)^{2}\left(2(\gamma_{11}-1)^{2}-\gamma_{12}^{2}\left(\sqrt{-\gamma_{11}^{2}+2\gamma_{11}-\gamma_{12}^{2}}-1\right)\right)}{2\left((\gamma_{11}-1)^{2}+\gamma_{12}^{2}\right)^{3}}\nonumber \\
 &  & +\frac{\left((\gamma_{11}-1)^{2}+\gamma_{12}^{2}\right)^{2}\left(4-\frac{\gamma_{12}^{2}\left(\gamma_{12}^{2}-1\right)}{\left(-\gamma_{11}^{2}+2\gamma_{11}-\gamma_{12}^{2}\right)^{3/2}}\right)-2\left((\gamma_{11}-1)^{2}+\gamma_{12}^{2}\right)\left(2(\gamma_{11}-1)^{2}-\gamma_{12}^{2}\left(\sqrt{-\gamma_{11}^{2}+2\gamma_{11}-\gamma_{12}^{2}}-1\right)\right)}{2\left((\gamma_{11}-1)^{2}+\gamma_{12}^{2}\right)^{3}}\label{eq:hess1}
\end{eqnarray}
}}{\footnotesize{
\begin{eqnarray}
\frac{\partial^{2}E}{\partial\gamma_{11}\partial\gamma_{12}} & = & -\frac{(\gamma_{11}-1)\gamma_{12}\left(-2\gamma_{11}^{6}+12\gamma_{11}^{5}+2\gamma_{11}^{2}\left(10\sqrt{-\gamma_{11}^{2}+2\gamma_{11}-\gamma_{12}^{2}}-9\gamma_{12}^{2}+1\right)-4\gamma_{11}\left(2\sqrt{-\gamma_{11}^{2}+2\gamma_{11}-\gamma_{12}^{2}}-3\gamma_{12}^{2}+1\right)\right)}{2\left(-\gamma_{11}^{2}+2\gamma_{11}-\gamma_{12}^{2}\right)^{3/2}\left(\gamma_{11}^{2}-2\gamma_{11}+\gamma_{12}^{2}+1\right)^{3}}\nonumber \\
 &  & -\frac{(\gamma_{11}-1)\gamma_{12}\left(\gamma_{12}^{2}\left(-2\gamma_{12}^{2}\left(2\sqrt{-\gamma_{11}^{2}+2\gamma_{11}-\gamma_{12}^{2}}+3\right)+4\sqrt{-\gamma_{11}^{2}+2\gamma_{11}-\gamma_{12}^{2}}+\gamma_{12}^{4}+1\right)\right)}{2\left(-\gamma_{11}^{2}+2\gamma_{11}-\gamma_{12}^{2}\right)^{3/2}\left(\gamma_{11}^{2}-2\gamma_{11}+\gamma_{12}^{2}+1\right)^{3}}\nonumber \\
 &  & -\frac{(\gamma_{11}-1)\gamma_{12}\left(+\gamma_{11}^{4}\left(4\left(\sqrt{-\gamma_{11}^{2}+2\gamma_{11}-\gamma_{12}^{2}}-6\right)-3\gamma_{12}^{2}\right)-4\gamma_{11}^{3}\left(4\sqrt{-\gamma_{11}^{2}+2\gamma_{11}-\gamma_{12}^{2}}-3\gamma_{12}^{2}-4\right)\right)}{2\left(-\gamma_{11}^{2}+2\gamma_{11}-\gamma_{12}^{2}\right)^{3/2}\left(\gamma_{11}^{2}-2\gamma_{11}+\gamma_{12}^{2}+1\right)^{3}}\label{eq:hess2}
\end{eqnarray}
}}{\footnotesize \par}

{\footnotesize{
\begin{eqnarray}
\frac{\partial^{2}E}{\partial^{2}\gamma_{12}^{2}} & = & \frac{-4\gamma_{12}^{2}\left((\gamma_{11}-1)^{2}+\gamma_{12}^{2}\right)\left(\frac{\gamma_{12}^{2}}{\sqrt{-\gamma_{11}^{2}+2\gamma_{11}-\gamma_{12}^{2}}}-2\sqrt{-\gamma_{11}^{2}+2\gamma_{11}-\gamma_{12}^{2}}+2\right)}{2\left((\gamma_{11}-1)^{2}+\gamma_{12}^{2}\right)^{3}}\nonumber \\
 &  & +\frac{8\gamma_{12}^{2}\left(2(\gamma_{11}-1)^{2}-\gamma_{12}^{2}\left(\sqrt{-\gamma_{11}^{2}+2\gamma_{11}-\gamma_{12}^{2}}-1\right)\right)-2\left((\gamma_{11}-1)^{2}+\gamma_{12}^{2}\right)\left(2(\gamma_{11}-1)^{2}-\gamma_{12}^{2}\left(\sqrt{-\gamma_{11}^{2}+2\gamma_{11}-\gamma_{12}^{2}}-1\right)\right)}{2\left((\gamma_{11}-1)^{2}+\gamma_{12}^{2}\right)^{3}}\nonumber \\
 &  & +\frac{\left((\gamma_{11}-1)^{2}+\gamma_{12}^{2}\right)^{2}\left(\frac{5\gamma_{12}^{2}}{\sqrt{-\gamma_{11}^{2}+2\gamma_{11}-\gamma_{12}^{2}}}-2\sqrt{-\gamma_{11}^{2}+2\gamma_{11}-\gamma_{12}^{2}}+\frac{\gamma_{12}^{4}}{\left(-\gamma_{11}^{2}+2\gamma_{11}-\gamma_{12}^{2}\right)^{3/2}}+2\right)}{2\left((\gamma_{11}-1)^{2}+\gamma_{12}^{2}\right)^{3}}\label{eq:hess3}
\end{eqnarray}
}}{\footnotesize \par}

\end{widetext}

\begin{figure}

\includegraphics[angle=-90,width=1\columnwidth]{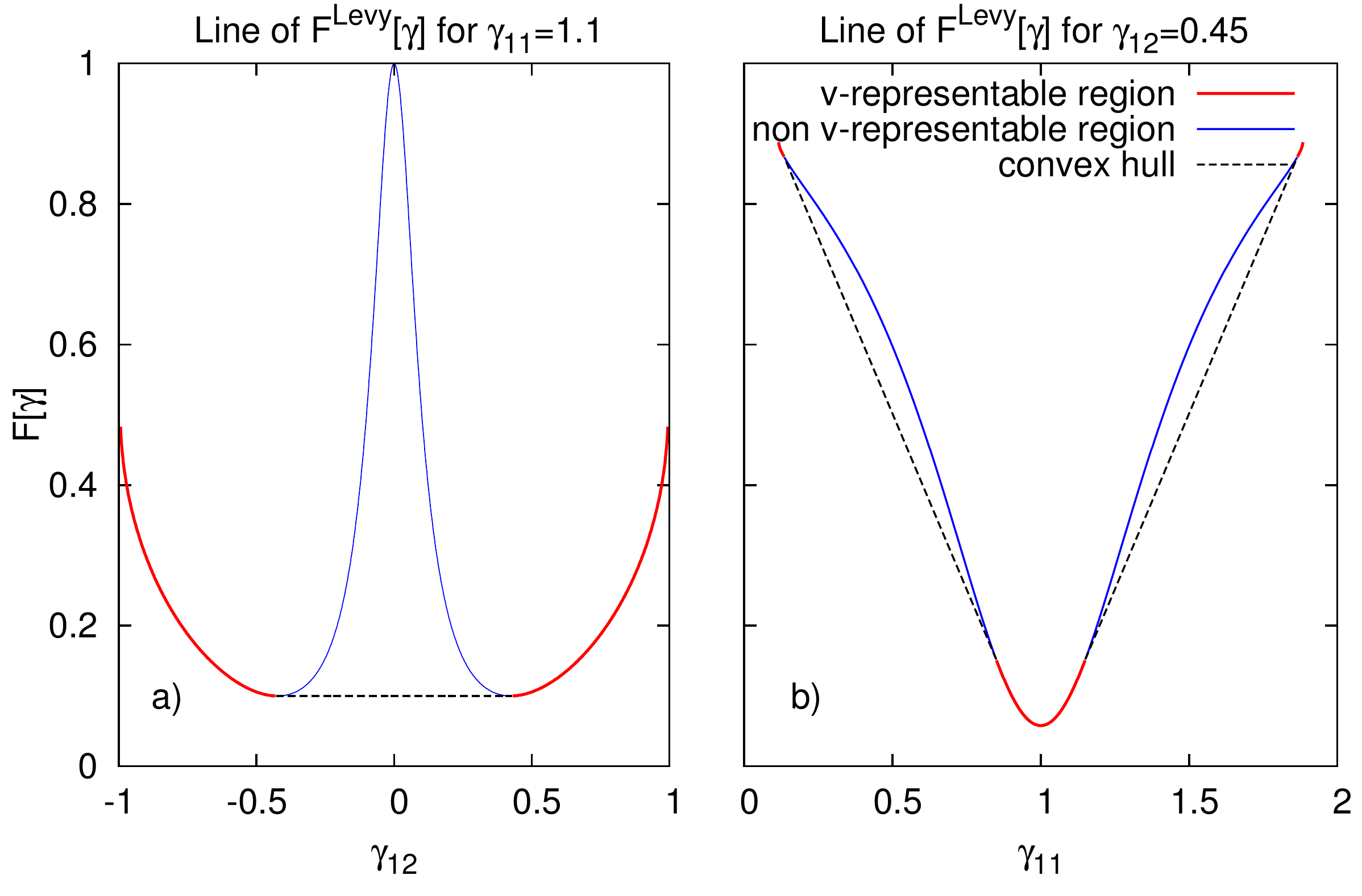}\caption{Two lines of $F^{{\rm Levy}}[\gamma]$ that illustrate non-$v$-representable
density matrices, due to the non convexity of the surface along the
given line.\label{fig:Two-lines-of} }

\end{figure}

For the discussion of $v$-representability, there are two common
counterexamples: the first is a one-electron density with a certain
type of cusp, given by Englisch and Englisch\cite{SuppEnglisch83253};
the other is a spherical $p$ density related to a degeneracy that
cannot be given by a single wavefunction\cite{SuppSavin96327}. The non-$v$-representable
density matrices shown here are very different to these two examples
and are only due to the nature of the energy surface of the exact
functional as shown in Fig. \ref{fig:Two-lines-of}.

\subsection{Derivation of Löwdin-Shull for Hubbard model}

Löwdin and Shull (LS) showed that the natural orbitals, $\phi_{k}$,
that diagonalize the density matrix and wavefunction for two electrons
are the same
\begin{eqnarray}
\Psi({\bf r},{\bf r}^{\prime}) & = & \sum_{k}c_{k}\phi_{k}({\bf r})\phi_{k}({\bf r}^{\prime})\\
\gamma({\bf r},{\bf r}^{\prime}) & = & \sum_{k}n_{k}\phi_{k}({\bf r})\phi_{k}({\bf r}^{\prime})
\end{eqnarray}
where $n_{k}=2c_{k}^{2}$. 
\begin{equation}
E^{{\rm LS}}[\Psi]=\sum_{i=1}^{2}c_{i}^{2}\left\{ 2h_{ii}+\langle ii|ii\rangle\right\} +2c_{1}c_{2}\langle11|22\rangle
\end{equation}
For two basis functions the minimum energy wavefunction comes from
the coefficients of $c_{1}$ and $c_{2}$ having opposite signs, $c_{1}=\sqrt{n_{1}/2}$
and $c_{2}=-\sqrt{n_{2}/2}$. Substituting this into the energy expression
for the wavefunction gives an expression in terms of the natural orbitals
and the natural orbital occupation numbers, $n_{k},$

\begin{equation}
F^{{\rm LS}}[\gamma]=\frac{1}{2}n_{a}\langle aa|aa\rangle+\frac{1}{2}n_{b}\langle bb|bb\rangle-\sqrt{n_{a}n_{b}}\langle aa|bb\rangle.\label{eq:Lowdin-1}
\end{equation}
There has been some recent interest in natural orbitals \cite{SuppGiesbertz13104110}
and natural orbital functionals that, for two electron systems, must
reduce to the Löwdin-Shull expression if they are to be exact, for
example the PNOF5 functional \cite{SuppPiris11164102,SuppPernal13127,SuppPiris13234109}. 

The eigenvalues of the density matrix $\gamma=\left(\begin{array}{cc}
\gamma_{11} & \gamma_{12}\\
\gamma_{12} & (2-\gamma_{11})
\end{array}\right)$ are
\begin{eqnarray}
(\gamma_{11}-n)((2-\gamma_{11}-n)-\gamma_{12}^{2} & = & 0\\
n^{2}-2n+2\gamma_{11}-\gamma_{11}^{2}-\gamma_{12}^{2} & = & 0
\end{eqnarray}

\begin{equation}
n_{\pm}=\left(-2\pm\sqrt{4-4(\gamma_{11}-1)^{2}+4\gamma_{12}^{2}}\right)/2
\end{equation}
\begin{eqnarray}
n_{\pm}=n_{a/b} & = & 1\pm\sqrt{(\gamma_{11}-1)^{2}+\gamma_{12}^{2}}.
\end{eqnarray}
The $\langle pp|qq\rangle$ integrals are in the natural orbital basis
and the coefficients of the natural orbitals ($C_{pi}$) are found
by substituting in the natural orbital numbers e.g. $\left(\gamma_{11}-n_{p}\right)C_{p1}+\gamma_{12}C_{p2}=0$
or (also using $C_{\pm i}=C_{(a/b)i}$) 
\begin{equation}
C_{\pm1}=\frac{(\gamma_{11}-1)\pm\sqrt{(\gamma_{11}-1)^{2}+\gamma_{12}^{2}}}{\gamma_{12}}C_{\pm2}\ \ {\rm and}\ \ C_{\pm1}^{2}+C_{\pm2}^{2}=1
\end{equation}
So overall, $C_{\pm1}^{2}=a_{\pm}^{2}/\left(\gamma_{12}^{2}+a_{\pm}^{2}\right)$
and $C_{\pm2}^{2}=\gamma_{12}^{2}/(a_{\pm}^{2}+\gamma_{12}^{2})$
and hence 
\begin{eqnarray}
F^{{\rm LS}} & = & \frac{1}{2}n_{a}(C_{a1}^{4}+C_{a2}^{4})U+\frac{1}{2}n_{b}(C_{b1}^{4}+C_{b2}^{4})U\nonumber \\
 &  & -\sqrt{n_{a}n_{b}}(C_{a1}^{2}C_{b1}^{2}+C_{a2}^{2}C_{b2}^{2})U.
\end{eqnarray}
For convenience, replace $r=(\gamma_{11}-1)$ and $S=\sqrt{r^{2}+\gamma_{12}^{2}}$,
to obtain the following expression\begin{widetext} 
\begin{eqnarray}
F^{{\rm LS}} & = & \frac{(1+S)}{2}\left[\left(\frac{(r+S)^{2}}{\gamma_{12}^{2}+(r+S)^{2}}\right)^{2}+\left(\frac{\gamma_{12}^{2}}{\gamma_{12}^{2}+(r+S)^{2}}\right)^{2}\right]+\frac{(1-S)}{2}\left[\left(\frac{(r-S)^{2}}{\gamma_{12}^{2}+(r-S)^{2}}\right)^{2}+\left(\frac{\gamma_{12}^{2}}{\gamma_{12}^{2}+(r-S)^{2}}\right)^{2}\right]\nonumber \\
 &  & \ \ \ \ \ \ \ \ \ -\sqrt{1-S^{2}}\left[\frac{(r+S)^{2}}{\gamma_{12}^{2}+(r+S)^{2}}\frac{(r-S)^{2}}{\gamma_{12}^{2}+(r-S)^{2}}+\frac{\gamma_{12}^{2}}{\gamma_{12}^{2}+(r+S)^{2}}\frac{\gamma_{12}^{2}}{\gamma_{12}^{2}+(r-S)^{2}}\right]
\end{eqnarray}
\end{widetext}This equation could be simplified further but we have
checked, by numerical evaluation with Fortran code, that it gives
identical results to Eq. (13).

\subsection{Complex }

\begin{figure}[h]
\includegraphics[angle=-90,width=1\columnwidth]{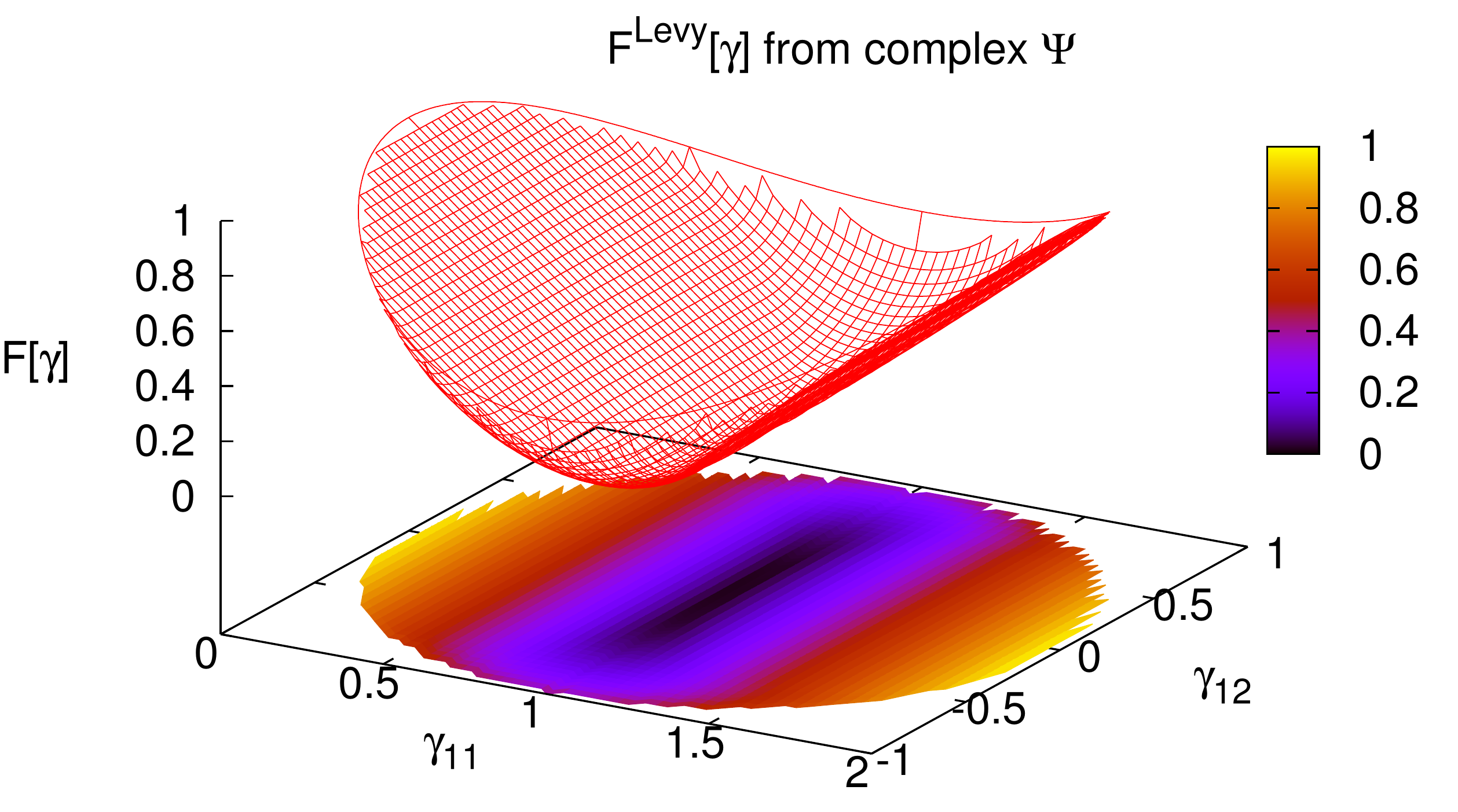}\caption{The exact functional allowing the wavefunction to be complex in the
Levy search.\label{fig:ComplexLevy}}
\end{figure}
The constrained search $\Psi\rightarrow\gamma$ can be expanded over
complex wavefunctions where the parameters, $a,b,c$, in the wavefunction 

\begin{eqnarray}
\Psi & = & \frac{a}{\sqrt{2}}\left[\mathcal{A}(\phi_{1}\alpha\phi_{2}\beta)+\mathcal{A}(\phi_{2}\alpha\phi_{1}\beta)\right]\nonumber \\
 &  & +b\mathcal{A}(\phi_{1}\alpha\phi_{1}\beta)+c\mathcal{A}(\phi_{2}\alpha\phi_{2}\beta)\label{eq:Wavefunction-1}
\end{eqnarray}
are allowed to be complex

\begin{eqnarray*}
a & = & a_{r}+ia_{i}\\
b & = & b_{r}+ib_{i}\\
c & = & c_{r}+ic_{i}
\end{eqnarray*}
In terms of these parameters there are the following constraints:
\begin{eqnarray*}
1 & = & a_{r}^{2}+a_{i}^{2}+b_{r}^{2}+b_{i}^{2}+c_{r}^{2}+c_{i}^{2}\\
\gamma_{11} & = & 2a_{r}^{2}+2a_{i}^{2}+b_{r}^{2}+b_{i}^{2}\\
\Re(\gamma_{12}) & = & \sqrt{2}(a_{r}b_{r}+a_{i}b_{i}+b_{r}c_{r}+b_{i}c_{i})
\end{eqnarray*}
The imaginary part $\Im(\gamma_{12})$ can be anything as it does
not enter the energy expression. A fourth constraint can be included
if the overall phase of the wavefunction is set to zero.

We now carry out a search over all possible wavefunctions minimizing
$E$ and a given $\gamma_{11}$ and $\Re(\gamma_{12})$, which gives
Fig. \ref{fig:ComplexLevy}. We do this by an explicit grid search
over the two remaining variables for each $\gamma_{11},\gamma_{12}$
that is specified. The resulting energy functional gives the same
result as the Hubbard expression Eq. (13) for all density matrices
except the non-$v$-representable set. For all possible FCI density
matrices it is, of course, in agreement with $F^{{\rm HK}}[\gamma_{v}]$.
For the non-$v$-representable set, $F_{{\rm complex}}^{{\rm Levy}}[\gamma]$
can be lower in energy, though this does not change any physics as
these points can never be minima of any Hamiltonian. In this case,
the functional numerically agrees with the ensemble functional considered
by Saubènere and Pastor\cite{SuppSaubanere1103511} given by a density
matrix that is an ensemble of two wavefunctions $\Gamma=a|\Psi_{a}\rangle\langle\Psi_{a}|+b|\Psi_{b}\rangle\langle\Psi_{b}|$.
It should be noted that when $F_{{\rm complex}}^{{\rm Levy}}[\gamma]$
is lower than Eq. (13) the solutions have a current and this may give
a connection to the exact functional in current DFT (CDFT) \cite{SuppVignale872360,SuppTellgren12062506}.

\subsection{Lieb maximization}

\begin{figure}[h]
\includegraphics[angle=-90,width=1\columnwidth]{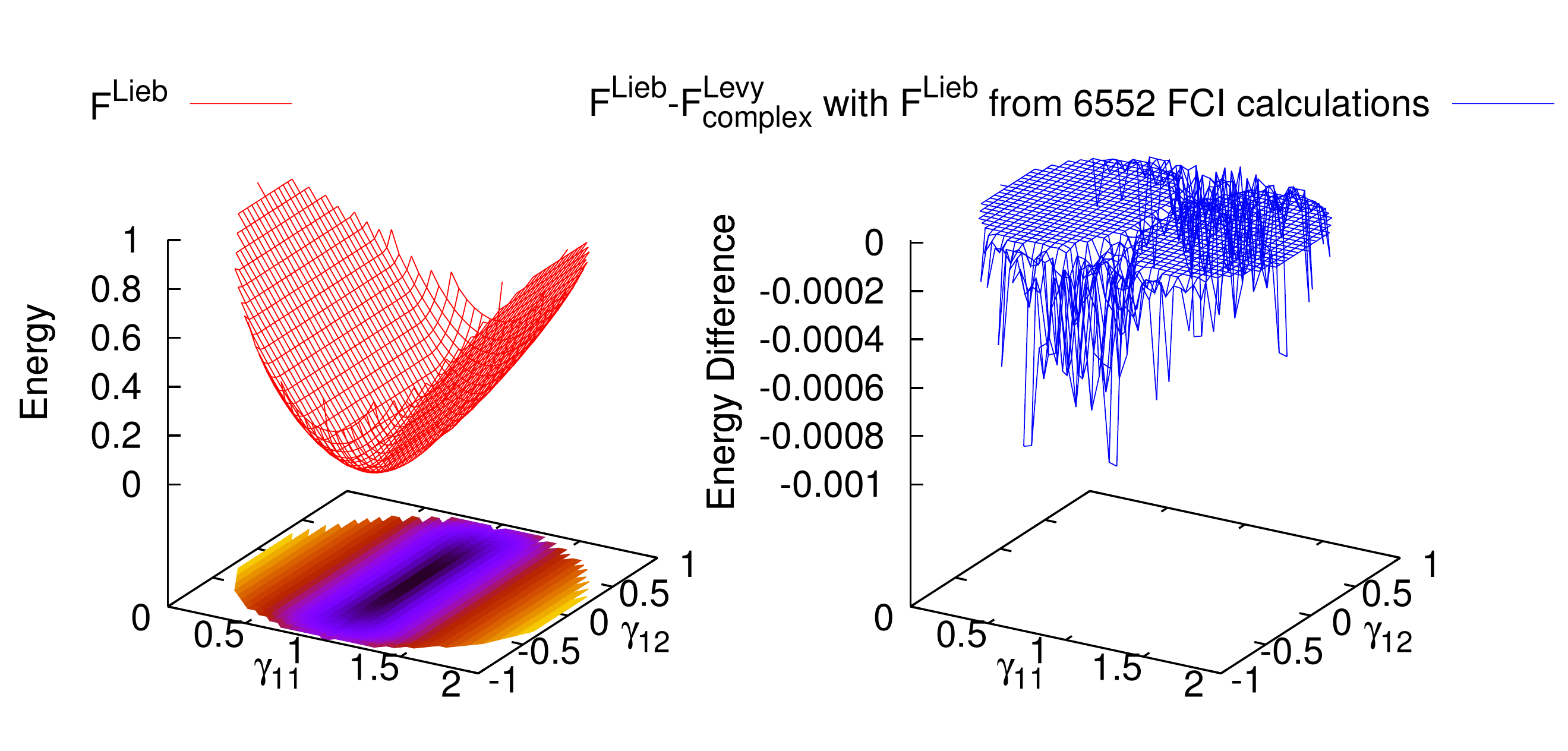}\caption{Functional $F^{{\rm Lieb}}[\rho]$ from Lieb maximization using 6552
FCI calculations\label{fig:Lieb Functional}}
\end{figure}

Another way to to calculate a bound for the functional is to perform
the Lieb maximization\cite{SuppLieb83243}, 
\begin{equation}
F^{{\rm Lieb}}[\gamma]=\sup_{v}\left\{ E_{v}-\gamma.v\right\} \label{eq:LiebMaximization}
\end{equation}
which is a supremum (a smallest upper bound which for any finite set
would just be a maximum) on the set of $v$. This means for a finite
set it would actually be a lower bound to the true minimum $F^{{\rm Lieb}}[\gamma]\le F^{{\rm Levy}}[\gamma]$.
The Lieb maximization is carried out using 6552 FCI calculations for
$v$, with $-10<t<10$ and $-10<\Delta\epsilon<10$. Over a grid of
density matrices, we compare directly with $F_{{\rm complex}}^{{\rm Levy}}$
from complex wavefunctions as in the region of non-$v$-representable
densities it is closest to the complex or ensemble form. Carrying
out the maximization $ $of Eq. (\ref{eq:LiebMaximization}) gives
the results in the left hand side of Fig. \ref{fig:Lieb Functional}
and the difference to $F_{{\rm complex}}^{{\rm Levy}}$ is shown in
the right-hand side. This difference is small and negative which illustrates
that the Lieb maximization only gives a lower bound to the true functional
that in this case is known exactly. Obviously, with more and more
FCI calculations $F^{{\rm Lieb}}$ would approach closer to the correct
result. The $F^{{\rm Lieb}}[\gamma]$ should not be used in minimizations
in the same way as $F^{{\rm Levy}}[\gamma]$ as it is a lower bound
rather than an upper bound. Finally it should be noted that $F^{{\rm Lieb}}[\gamma]$
is everywhere convex by construction and cannot, for example, contribute
to the discussion on $v$-representability.

\subsection{Approximate Density Matrix Functionals}

We consider various approximate density matrix functionals including
Hartree-Fock as a density matrix functional, Muller\cite{SuppMuller84446},
Power \cite{SuppSharma08201103}. Here the value of the natural orbital
occupation numbers $0\le n_{i}\le2$ and the two-electron integrals
$\langle pq|rs\rangle=\int\int\phi_{p}^{*}({\bf r})\phi_{r}({\bf r})V_{ee}({\bf r},{\bf r}^{\prime})\phi_{q}^{*}({\bf r}^{\prime})\phi_{s}({\bf r}^{\prime}){\rm d}{\bf r}{\rm d}{\bf r}^{\prime}$
which in the asymmetric two-site Hubbard model just work out to be
$\langle pq|rs\rangle=\sum_{i=1,2}C_{pi}C_{qi}C_{ri}C_{si}$ in terms
of the orbitals coefficients $C_{pi}$ ($|p\rangle=\sum_{i=1,2}C_{pi}c_{i}^{\dagger}|{\rm vac}\rangle$)

\[
F^{{\rm Hartree-Fock}}=\frac{1}{2}n_{i}n_{j}\langle ij|ij\rangle-\frac{1}{4}n_{i}n_{j}\langle ii|jj\rangle
\]
\[
F^{{\rm M\ddot{u}ller}}=\frac{1}{2}n_{i}n_{j}\langle ij|ij\rangle-\frac{1}{2}\sqrt{n_{i}n_{j}}\langle ii|jj\rangle
\]
\[
F^{{\rm Power}}=\frac{1}{2}n_{i}n_{j}\langle ij|ij\rangle-\frac{1}{2}(n_{i}n_{j})^{\alpha}\langle ii|jj\rangle
\]
In the paper we use a value $\alpha=0.675$ that has recently been
used for Mott insulators

\subsection{Gutzwiller approximate wavefunction}

The Gutzwiller wavefunction \cite{SuppGutzwiller63159} is a parametrized
wavefunction of the form
\begin{eqnarray*}
\Psi & = & \frac{1}{\sqrt{2}}\left[\mathcal{A}(\phi_{1}\alpha\phi_{2}\beta)+\mathcal{A}(\phi_{2}\alpha\phi_{1}\beta)\right]\\
 &  & +g\left[\mathcal{A}(\phi_{1}\alpha\phi_{1}\beta)+\mathcal{A}(\phi_{2}\alpha\phi_{2}\beta)\right]
\end{eqnarray*}
When $g=1$ it is the Hartree-Fock wavefunction for orbitals $\phi=\frac{1}{\sqrt{2}}(\phi_{1}+\phi_{2})$.
The basic idea is that in an H$_{2}$ like system as $g\rightarrow0$
it goes to the Heitler-London wavefunction. In the asymmetric two-site
Hubbard model we consider an orbital of the form $\phi=c_{1}\phi_{1}+\sqrt{1-c_{1}^{2}}\phi_{2}$
and a Gutzwiller wavefunction
\begin{figure}[!ht]
\includegraphics[angle=-90,width=1\columnwidth]{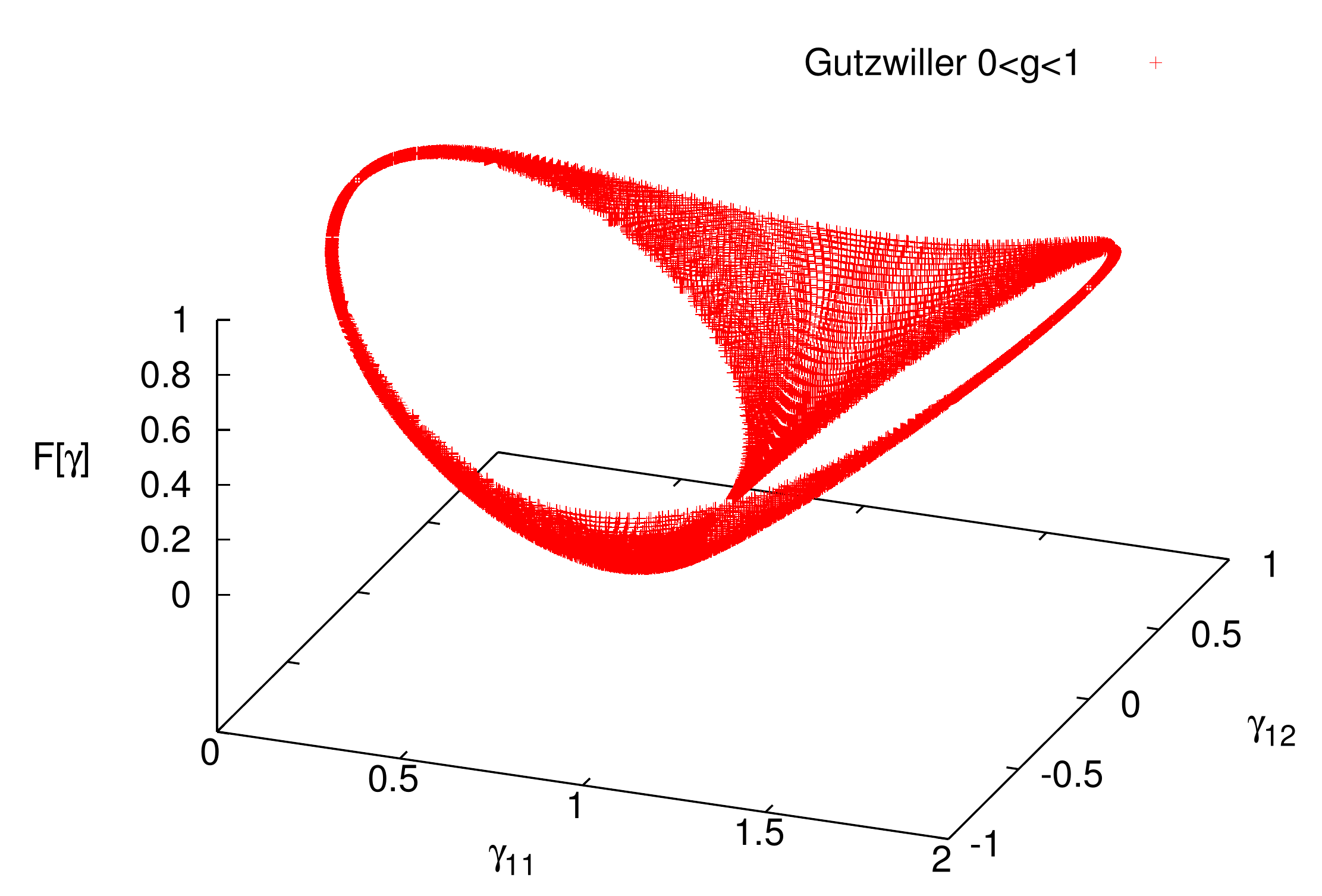}\caption{}
\end{figure}
\begin{eqnarray*}
\Psi^{{\rm GWA}} & = & \frac{2c_{1}\sqrt{1-c_{1}^{2}}}{\sqrt{2}}\left[\mathcal{A}(\phi_{1}\alpha\phi_{2}\beta)+\mathcal{A}(\phi_{2}\alpha\phi_{1}\beta)\right]\\
 &  & +g\left[c_{1}^{2}\mathcal{A}(\phi_{1}\alpha\phi_{1}\beta)+(1-c_{1}^{2})\mathcal{A}(\phi_{2}\alpha\phi_{2}\beta)\right]
\end{eqnarray*}
If we consider all possible values of $c_{1}$ and $-1\le g\le1$
we get the following density matrices and 
\[
F[\Psi^{{\rm GWA}}]=\frac{\langle\Psi^{{\rm GWA}}|V_{ee}|\Psi^{{\rm GWA}}\rangle}{\langle\Psi^{{\rm GWA}}|\Psi^{{\rm GWA}}\rangle}.
\]
For other values of $|g|>1$ the wavefunction is no longer a ground
state wavefunction.

\subsection{Functional for $N=0,1,2,3$ and $4$}

The functional is calculated for different integer numbers of electrons
($N=0,1,2,3$ and 4), where the trace of the density matrix $\gamma_{11}+\gamma_{22}=N$.
At $N=0$, $F[\gamma]=0$ and there is only one allowed density matrix
$\gamma_{11}=\gamma_{12}=0$. For $N=1$, $F[\gamma]=0$ as there
is no electron-electron interaction, however, the allowable density
matrices from a pure state wavefunction are now defined by a circle
$\gamma_{12}=\sqrt{\left(\gamma_{11}-0.5\right)^{2}-0.5^{2}}.$ Inside
this circle are ensemble-$N$-representable density matrices but they
cannot come from a pure-state wavefunction. For $N=2$, $F[\gamma]$
is that of Eq. (13). For $N=3$, $F[\gamma]=1$ at the allowed pure-state
density matrices defined by a different circle $\gamma_{12}=\sqrt{\left(\gamma_{11}-1.5\right)^{2}-0.5^{2}}.$
Also at $N=4$, $F[\gamma]=2$ at density matrix $\gamma_{11}=2$,
$\gamma_{12}=0$. All these integer parts of the exact functional
are pictured in the supplementary information.\ref{fig:N=00003D01234}

\begin{figure}[!h]
\includegraphics[width=0.5\textwidth]{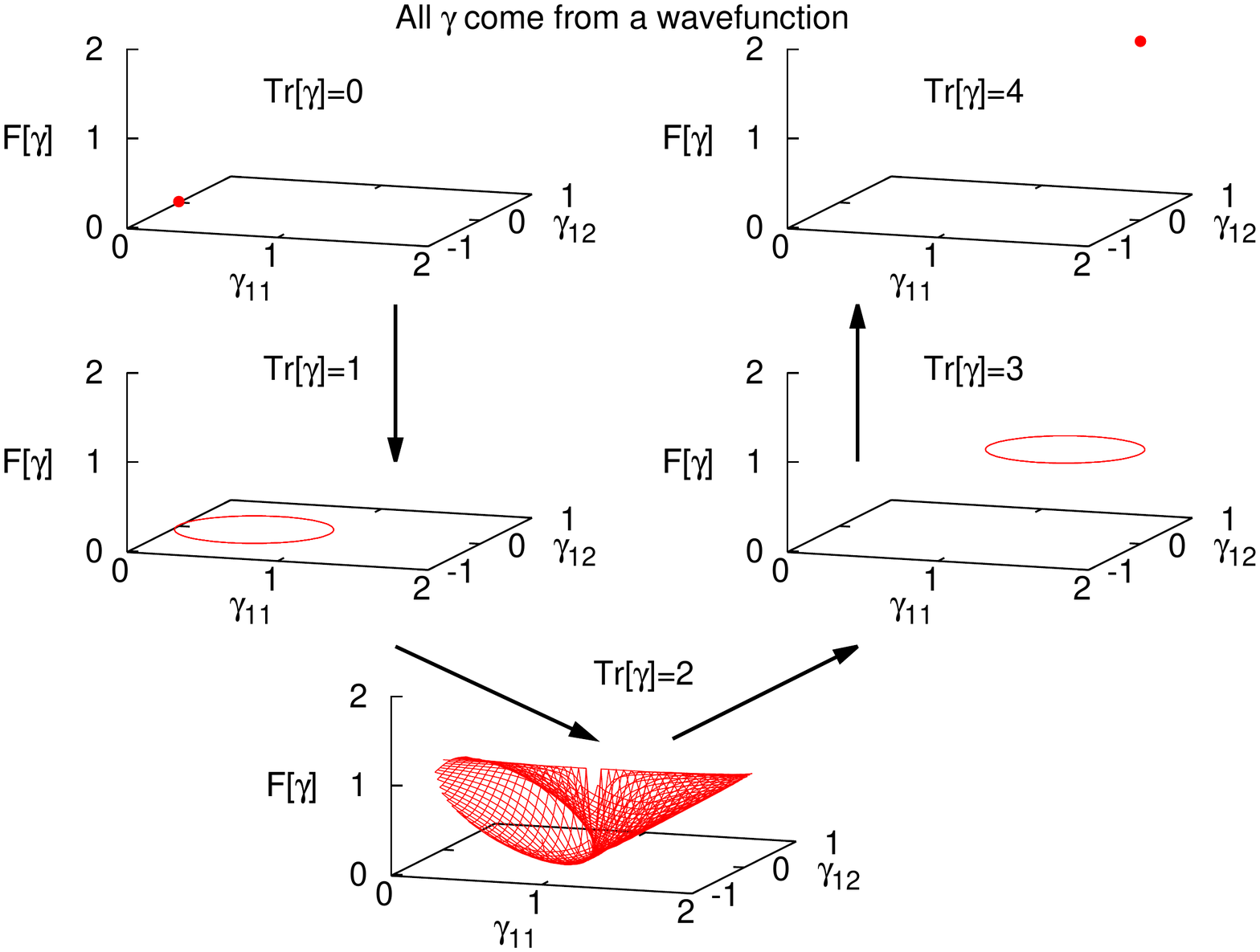}\caption{$F[\gamma]$ for $N=0,1,2,3,4$ electrons\label{fig:N=00003D01234}}
\end{figure}


%
%
%

\subsection{Other ensembles for $N=1.5$ electrons}
In the consideration of fractional numbers of electrons the argument
of convexity of $E$ vs $N$ is often used to simplify the ensembles
that have to be taken. 
\begin{figure}[!ht]
\includegraphics[angle=-90,width=0.5\textwidth]{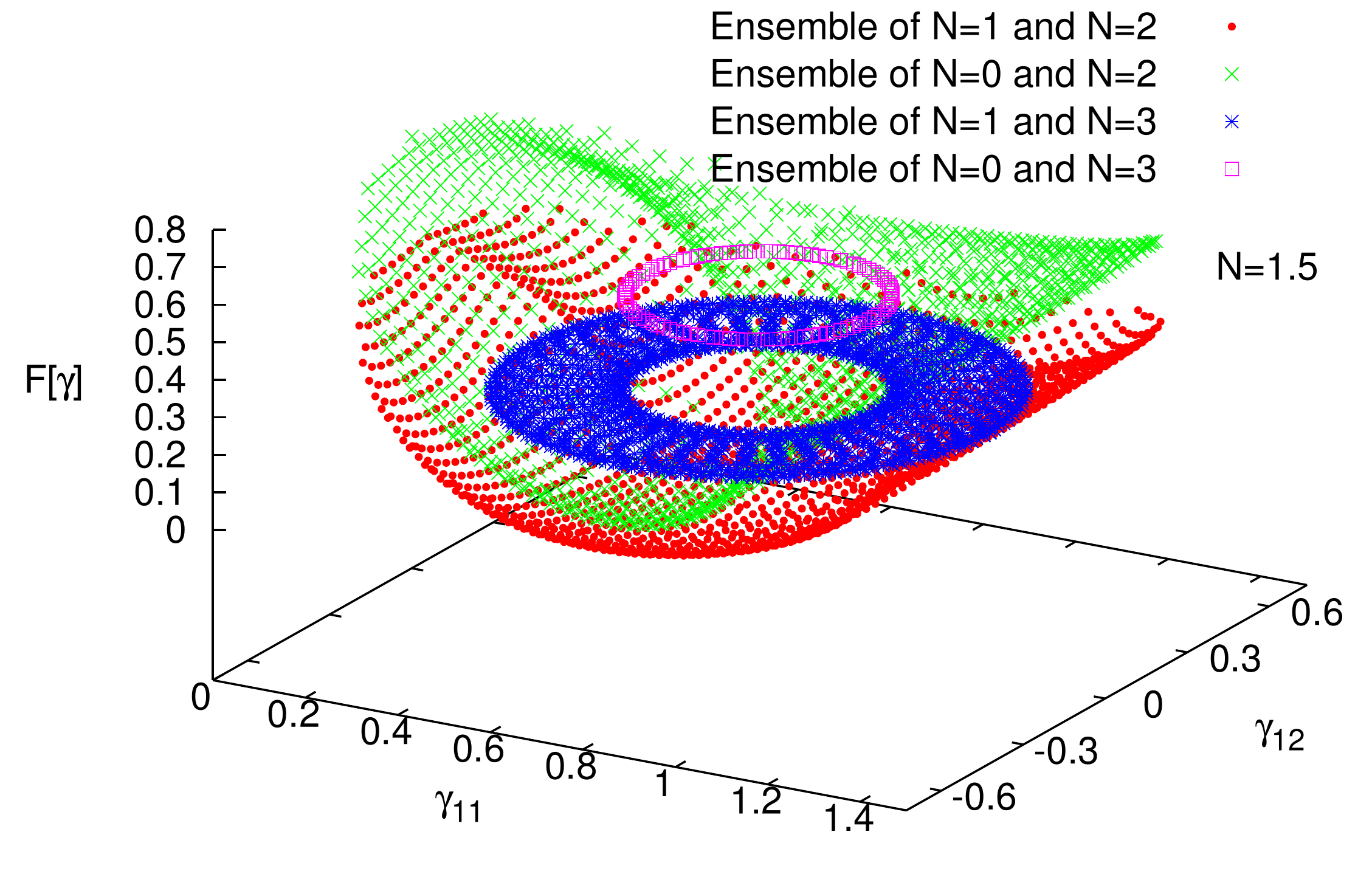}

\caption{Different ensemble formation of $F[\gamma^{N=1.5}]$ combining pure
state wavefunction for $N=0,1,2,3$ electrons \label{fig:Different-ensemble}}
\end{figure}
If convexity is true, the lowest energy ensemble will always be given
by the combination of the two integers at either side, e.g. $\Gamma_{N+\delta}=(1-\delta)\Gamma_{N}+\delta\Gamma_{N+1}$.
However, convexity has not been proven, with definitely known counterexamples
for certain electron-electron interactions, which indicates that most
certainly convexity is not a general property of Hamiltonians \cite{SuppLieb83243}.
Here, we test convexity for the two-site Hubbard hamiltonians, by
taking ensembles of different electron numbers. We consider different
pair-wise ensembles $\Gamma_{N=1.5}=a|\Psi_{n_{1}}\rangle\langle\Psi_{n_{1}}|+b|\Psi_{n_{2}}\rangle\langle\Psi_{n_{2}}|$
with $\{n_{1},n_{2}\}=\{1,2\},\{0,2\},\{1,3\},\{0,3\}$. We have also
considered all possible ensembles, including those of three and four
different particle numbers up to $N=4$, all of these lie higher in
energy.

\bibliographystyle{aipnum4-1}

\subsection{Supplementary Animations}
Supplementary animated gifs can be found in the arXiv source file or currently available via the following hyperlinks \\
\\
1) \ \ \href{http://people.ds.cam.ac.uk/ajc54/gifs/Vary_t.gif}{\underline{\color{blue}Varying $t$ with fixed $\Delta \epsilon$}}\\
2) \ \ \href{http://people.ds.cam.ac.uk/ajc54/gifs/StrongCorrelationVary_DeltaEpsilon.gif}{\underline{\color{blue}Electron transfer by varying $\Delta \epsilon$ for $t=1.0,0.2,0.05$}}\\
3) \ \ \href{http://people.ds.cam.ac.uk/ajc54/gifs/FractionalVary_t.gif}{\underline{\color{blue}Varying $t$ for fractional number of electrons, $N$}}

\end{document}